\documentclass{Interspeech}

\interspeechcameraready

\title{CMT-LLM: Contextual Multi-Talker ASR Utilizing Large Language Models}

\author[affiliation={1,2}]{Jiajun}{He}
\author[affiliation={2}]{Naoki}{Sawada}
\author[affiliation={2}]{Koichi}{Miyazaki}
\author[affiliation={3}]{Tomoki}{Toda}

\affiliation{Graduate School of Informatics}{Nagoya University}{Japan}
\affiliation{AI Lab}{CyberAgent}{Japan}
\affiliation{Information Technology Center}{Nagoya University}{Japan}
\email{\href{mailto:jiajun.he@g.sp.m.is.nagoya-u.ac.jp}{jiajun.he@g.sp.m.is.nagoya-u.ac.jp}, sawada\_naoki@cyberagent.co.jp, miyazaki\_koichi\_xa@cyberagent.co.jp, tomoki@icts.nagoya-u.ac.jp}
\keywords{Multi-talker ASR, Contextual ASR, Large Language Models, Serialized Output Training, Contextual Biasing}

\usepackage{comment}
\usepackage{multirow, threeparttable, booktabs, makecell}
\usepackage[table]{xcolor}
\usepackage{adjustbox}
\usepackage{hyperref}
\usepackage{CJKutf8}
\usepackage{footmisc} 

\begin{document}
\maketitle

\let\thefootnote\relax
\footnotetext{This work was done during Jiajun He's internship at CyberAgent.}

\begin{abstract}

In real-world applications, automatic speech recognition (ASR) systems must handle overlapping speech from multiple speakers and recognize rare words like technical terms. Traditional methods address multi-talker ASR and contextual biasing separately, limiting performance in complex scenarios. We propose a unified framework that combines multi-talker overlapping speech recognition and contextual biasing into a single task. Our ASR method integrates pretrained speech encoders and large language models (LLMs), using optimized finetuning strategies. We also introduce a two-stage filtering algorithm to efficiently identify relevant rare words from large biasing lists and incorporate them into the LLM’s prompt input, enhancing rare word recognition.Experiments show that our approach outperforms traditional contextual biasing methods, achieving a WER of 7.9\% on LibriMix and 32.9\% on AMI SDM when the biasing size is 1,000, demonstrating its effectiveness in complex speech scenarios.

\end{abstract}

\section{Introduction}

Multi-talker automatic speech recognition (ASR), particularly in overlapping speech scenarios, remains a major challenge. Existing methods include permutation invariant training \cite{kolbaek2017multitalker}, heuristic error assignment \cite{lu2021streaming}, and serialized output training (SOT) \cite{shi2024serialized}. Among these, SOT has gained attention for resolving speaker arrangement uncertainty by concatenating transcriptions in speech order. However, it relies heavily on long-context modeling, where attention-based encoder-decoder (AED) models struggle with inter-speaker dependencies.

Another key challenge in ASR is recognizing rare words, such as proper names and technical terms, which are underrepresented in training data \cite{he2023enhancing}. Contextual biasing methods address this by incorporating external biasing lists, but existing approaches have limitations. Shallow fusion adjusts decoding scores but struggles with large lists and dynamic contexts \cite{le2021deep}. Deep biasing \cite{10502142} improves accuracy by integrating biasing features but requires retraining and architectural 
modifications. Deep context models \cite{huang2023contextualized} leverage contextual text encoders but are computationally intensive. Contextual ASR error correction \cite{he2023ed, he2024mf, he2025pm} refines outputs post-recognition but depends on initial ASR hypotheses, introducing latency.  

In real-world applications, multi-talker ASR and contextual biasing often arise as intertwined challenges. For example, in meeting transcription, the ASR system needs to both distinguish between speakers and accurately recognize domain-specific terms and proper nouns. Similarly, in customer service settings with concurrent conversations, the system must dynamically identify each speaker's contributions while incorporating contextual information to enhance decoding accuracy. Despite their inherent interconnection, existing research typically treats multi-talker ASR and contextual biasing as separate tasks. To bridge this gap, we propose a novel task: the integrated application of contextual biasing within multi-talker ASR, aimed at improving the recognition of rare words in overlapping speech scenarios. To the best of our knowledge, this is the first study to combine multi-talker ASR with contextual biasing.

In recent years, the application of large language models (LLMs) in speech processing has garnered increasing attention \cite{bai2024seed, ma2024embarrassingly, meng2024large, shi2024advancing, yang2024mala}. With their remarkable global modeling capabilities, LLMs can effectively capture inter-speaker dependencies in multi-talker scenarios, producing semantically coherent and natural transcription outputs \cite{meng2024large, shi2024advancing}. In the context of contextual biasing, LLMs leverage prompt-based learning to flexibly integrate dynamic biasing lists provided by users, enabling accurate recognition of rare words while significantly reducing reliance on complex decoding mechanisms \cite{yang2024mala}. Consequently, LLMs demonstrate unique advantages in addressing the dual challenges of multi-talker ASR and contextual biasing.

\begin{figure*}[t]
  \centering
  \includegraphics[width=2\columnwidth]{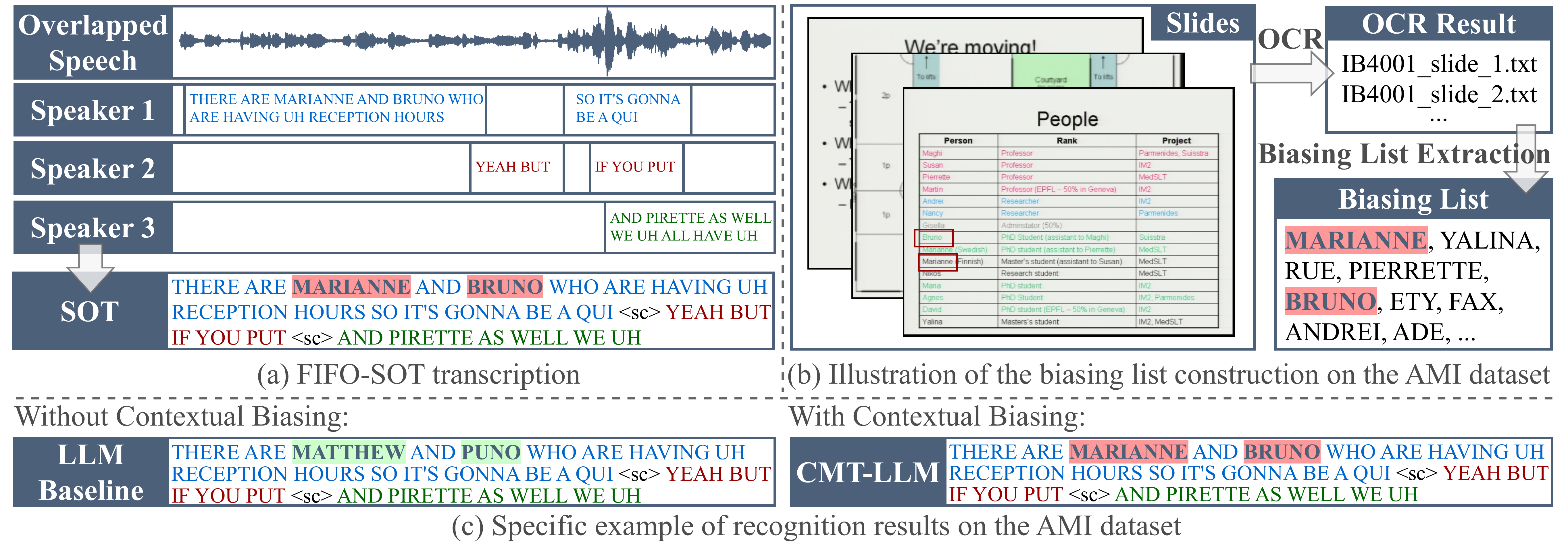}
  \vspace{-4mm}
  \caption{Illustration of the visual-grounded contextual multi-talker ASR pipeline.}
  \vspace{-4mm}
  \label{fig:pipeline}
\end{figure*}

To achieve this, we propose a comprehensive framework integrating a pretrained speech encoder, projector, and LLM. 
Specifically, we adopt a two-stage finetuning strategy to address the challenges posed by the multi-talker ASR task and the incorporation of large biasing lists in a practical setting:
first, the self-supervised learning (SSL) speech encoder is finetuned using the conventional SOT method. Next, we freeze the finetuned SSL speech encoder and LLM, training only the projector while efficiently finetuning the LLM using low-rank adaptation (LoRA). Moreover, considering the challenges posed by large biasing lists (e.g., thousands of words) in practical applications—where LLM’s prompt-based learning cannot effectively handle such extensive vocabularies—we filter and select the most relevant rare words based on coarse decoding results from the first stage. These refined small-scale biasing lists are then incorporated into the LLM’s prompt input, effectively addressing the challenges of large biasing lists and significantly improving contextual biasing performance, ultimately improving ASR accuracy.

\begin{figure}[!t]
  \centering
  \includegraphics[width=1\columnwidth]{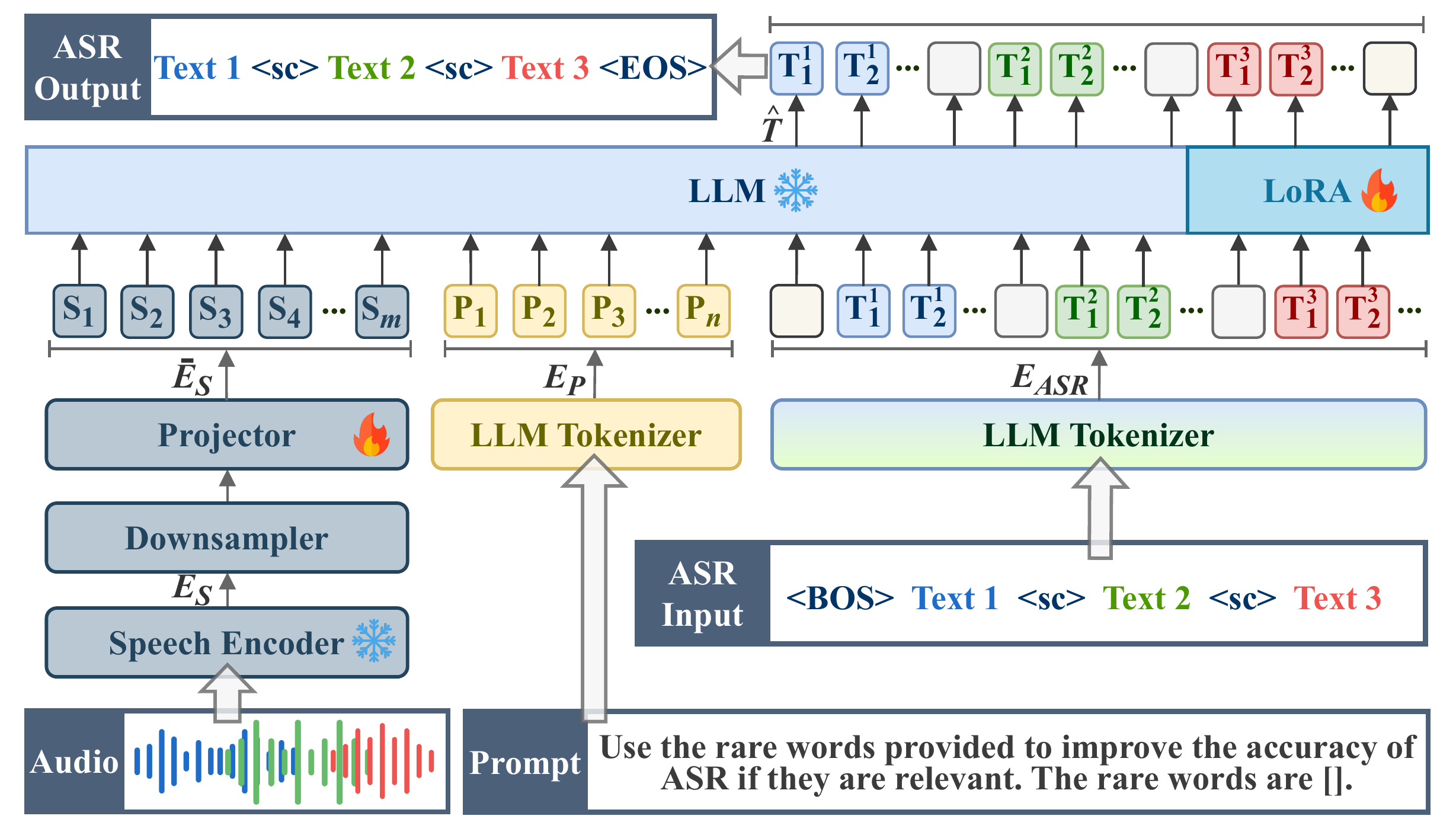}
  \vspace{-6mm}
  \caption{Overall architecture of the CMT-LLM model.}
  \label{fig:model}
  \vspace{-7mm}
\end{figure}

\section{Proposed Method}

\subsection{Problem Formulation}
\label{section2.0}
The contextual multi-talker ASR problem can be formalized as the mapping function $f(S, C) = T$, where the speech $S = (S_1, S_2, \cdots , S_M)$  contains $M$ acoustic frames, the context $C = (C_1, C_2, \cdots , C_L) \in \mathbb{R}^{L}$ denotes the biasing list containing $L$ rare words, and the ground truth transcript $T = (T_1^1, T_2^1, \cdots , \texttt{<sc>} , T_1^2, T_2^2, \cdots, \texttt{<sc>} , T_1^3, T_2^3, \cdots)  \in \mathbb{R}^{N}$ is the token sequence. In $T$, $T_a^b$ represents the $a$-th token of the $b$-th speaker and $N$ denotes the length of the transcript.  Following previous works \cite{kanda2020serialized, shi2024serialized}, we use the SOT method to address the multi-talker ASR problem. Specifically, assuming the number of speakers is 3, the transcriptions of different speakers are concatenated by inserting the speaker change symbol “\texttt{<sc>}”, creating the reference transcription for overlapping speech. The concatenation order follows the speaking time of each speaker, known as first-in first-out (FIFO), as shown in Fig. \ref{fig:pipeline}(a).

\vspace{-3mm}
\subsection{Proposed CMT-LLM}
In this section, we introduce a contextual multi-talker ASR method with LLM (\textbf{CMT-LLM}). As illustrated in Fig. \ref{fig:model}, like most previous LLM studies \cite{ma2024embarrassingly, meng2024large, shi2024advancing,   yang2024mala, hu-etal-2024-wavllm}, the proposed architecture comprises three key components: a speech encoder, a linear projector, and an LLM decoder.
First, the speech encoder processes the overlapping speech signal and extracts the corresponding speech representation $E_S \in \mathbb{R}^{M \times d_S}$, where $d_S$ represents the dimension of the extracted speech features. To make the representation more manageable for the LLM, we apply a 1D convolutional layer for downsampling, reducing the temporal resolution by a factor of $n$. The resulting features are then passed through a projection module consisting of two linear layers, transforming them into a speech embedding $\bar{E}_S \in \mathbb{R}^{\frac{M}{n} \times d_P}$. Here, $d_P$ denotes the dimension of the projected speech embedding, which matches the hidden size of the LLM to ensure compatibility with its input format.


For contextual multi-talker ASR, we enhance the model’s performance by incorporating a biasing list of rare words into the LLM’s prompt, enabling more accurate recognition of infrequent and potentially error-prone vocabulary. This prompt, which includes task-specific instructions and supplementary contextual information, undergoes tokenization and encoding to produce the prompt text embedding $E_P \in \mathbb{R}^{P \times d_P}$, where $P$ represents the length of the prompt. The process for generating the biasing list is detailed in Section \ref{sec:3.2}.


During training, the SOT-style multi-talker transcript is tokenized to obtain the ASR embedding $E_{ASR} \in \mathbb{R}^{N \times d_P}$. This embedding, together with the speech embedding $\bar{E}_{S}$ and the prompt text embedding $E_P$, is provided as input to the LLM, which is trained to predict the target transcript $\hat{T} \in \mathbb{R}^{N}$.
The model’s performance is optimized by minimizing the cross-entropy loss between the predicted transcript $\hat{T}$ and the ground truth $T$. 
During inference, both the speech embedding and the prompt text embedding (including task-specific instructions and the biasing list) are provided to the LLM, which then generates the ASR transcript in an autoregressive manner. To evaluate the impact of contextual information, we also assess a variant of the model without the biasing list in the prompt, referred to as the LLM Baseline.

The biasing list size in the LLM prompt is limited to 100 words to minimize computational costs during training. The detailed construction process will be described in Section \ref{sec:3.2}. However, in real-world applications, the biasing list may contain tens of thousands of words in the inference stage, which challenges the CMT-LLM’s ability to select the most relevant words effectively, causing significant performance degradation. 

To address this issue, we introduce a two-stage filtering approach that targets the challenges of large biasing lists inspired by previous studies \cite{yang23o_interspeech, huang23d_interspeech, xu23d_interspeech, yang2024ctc} during inference.
Specifically, in the first stage, we finetune a SSL pretrained speech model on the target dataset, adding a simple CTC head. A greedy decoding algorithm is then employed to generate initial predictions with lower computational overhead. The aims of this step are: 1) Previous research indicates that finetuning a pretrained speech encoder using traditional methods before applying LLM-based ASR training typically outperforms using the pretrained speech encoder directly; 2) The initial decoding results help filter out irrelevant rare words from the large biasing list.

In the second stage, we process the initial CTC decoding results by first removing the most common 5,000 words to retain more distinctive rare words. For example, if the ground truth transcription is “… MORE THAN THE SPEAKER CHARACTERISATION AS M STEVE …” and the initial decoding is “… MORE THAN THE SPEAKER CHARACE THSATION AS STEE …,” after removing common words, the remaining words are “CHARACE THSATION” and “STEE”. Next, we generate all possible subcombinations of these remaining words to more thoroughly select the most relevant terms. For example, given “CHARACE THSATION”, the possible segments include “CHARACE”, “THSATION”, and “CHARACE THSATION”. We calculate the word-based edit distance between these segments and the words in the biasing list, selecting the closest match, “CHARACTERISATION”. We notice that phoneme-based edit distance calculation and text semantic similarity calculation are both slower and less effective than the simplest word-based edit distance calculation. If only the individual words “CHARACE” and “THSATION” are considered, the target word “CHARACTERISATION” might be overlooked. 
To balance accuracy, efficiency, and computational cost, we match each candidate word with its Top-10 related words during filtering. 
Finally, we merge the selected relevant rare words, remove duplicates, and add them to the LLM prompt.


\vspace{-3mm}
\section{Experimental Evaluation}

\subsection{Implementation Details}
\label{ssec:Implementation Details}

Our method was trained on 4 NVIDIA A100 80 GB GPUs and the batch size was set to 2. We employed the finetuned \texttt{WavLM-Large}\footnote{\href{https://huggingface.co/microsoft/wavlm-large}{https://huggingface.co/microsoft/wavlm-large}} \cite{chen2022wavlm} as the speech encoder, processing 16 kHz sampled audio into feature embeddings with a frame rate of 50 Hz and a dimension of 1,024. These embeddings underwent downsampling ($n$ = 5) and were transformed via two linear projection layers. This process produced speech embeddings with a final frame rate of 10 Hz and a dimension of 4,096.

For the LLM module, we integrated \texttt{Vicuna-7B}\footnote{\href{https://huggingface.co/lmsys/vicuna-7b-v1.5}{https://huggingface.co/lmsys/vicuna-7b-v1.5}} \cite{chiang2023vicuna}, a variant of \texttt{LLaMA} \cite{touvron2023llama} finetuned on ShareGPT conversational data. During training, only the projection layer was trained, while the speech encoder and LLM remained frozen. We adopted the AdamW optimizer \cite{loshchilov2017decoupled} with a learning rate of 0.0001, hyperparameters $\beta$ set to (0.9, 0.999), epsilon to 1e-08, and weight decay to 1e-6. The training strategy followed a linear warmup schedule with 1,000 warmup steps, continuing for up to 100,000 steps, with early stopping triggered by stagnating validation loss. Additionally, LoRA was applied for LLM finetuning, with $\alpha$ set to 32, $r$ set to 8, and dropout set to 0.05.
We defined specific prompt formats for different models. The prompt for the LLM Baseline was simply: “Transcribe speech to text.”. In contrast, the CMT-LLM employed a more detailed prompt with contextual information: “Use the rare words provided to improve the accuracy of ASR if they are relevant. The rare words are [...].”, where the biasing word list is dynamically inserted into the brackets. During inference, we utilized beam search decoding with a beam size of 4.

\begin{table}[ht]
\centering
\vspace{-2mm}
\caption{WER performance comparison with different multi-talker ASR models on LibriMix (\%) with 1,000 distractors.}
\vspace{-3mm}
\label{tab:per_comparison}
\begin{tabular}{l|c|cccc}
\toprule
\multirow{2}{*}{\textbf{Model}} & \multirow{2}{*}{\textbf{Year}} & \multicolumn{2}{c}{\textbf{LibriMix}} \\
\cmidrule(lr){3-4}
& & \textbf{Dev} & \textbf{Test} \\
\midrule
\midrule
 Conditional-Conformer-CTC \cite{guo21_interspeech} & 2021 & 24.5 & 24.9 \\
WavLM-CTC \cite{chen2022wavlm} & 2022 & 23.0 & 20.3 \\
Whisper (Small)\footnotemark[3] & 2023 & 26.0 & 25.0 \\
Conformer\footnotemark[3] & 2022 & 24.7 & 23.3 \\
\hspace{1em} + WavLM-Large upstream\footnotemark[3] & 2023 & 19.4 & 17.1 \\
GEncSep \cite{shi2024serialized} & 2024 & 17.2 & 15.0 \\
\midrule
LLM Baseline & 2025 & 12.7 & 9.2 \\
\rowcolor{gray!20}
\textbf{CMT-LLM (ours)} & \textbf{2025} & \textbf{8.1} & \textbf{7.3} \\
\bottomrule
\end{tabular}
\vspace{-1mm}
\end{table}


\begin{table}[ht]
\centering
\vspace{-1mm}
\caption{WER performance comparison with different multi-talker ASR models on AMI (\%) with 1,000 distractors.}
\vspace{-3mm}
\label{tab:amiper_comparison}
\begin{adjustbox}{max width=0.47\textwidth}

\begin{tabular}{l|c|cccccccc}
\toprule
\multirow{3}{*}{\textbf{Model}} & \multirow{3}{*}{\textbf{Year}} & 
\multicolumn{6}{c}{\textbf{AMI}} \\
& &
\multicolumn{2}{c}{\textbf{IHM-Mix}} & \multicolumn{2}{c}{\textbf{SDM}} & \multicolumn{2}{c}{\textbf{MDM}}\\
\cmidrule(lr){3-8}
& & \textbf{Dev} & \textbf{Test} & \textbf{Dev} & \textbf{Test} & \textbf{Dev} & \textbf{Test}\\
\midrule
\midrule
WavLM-CTC \cite{chen2022wavlm} & 2022 & 34.4 & 34.3 & 39.8 & 44.0 & 38.1 & 41.5\\
SURT 2.0 (Large) \cite{10.1109/TASLP.2023.3318398} & 2023 & - & 36.8 & - & 62.5 & - & 44.4\\
\hspace{1em} + Adaptation \cite{10.1109/TASLP.2023.3318398} & 2023 & - & 35.1 & - & 44.6 & - & 41.4\\
\midrule
LLM Baseline & 2025 & 24.1 & 23.5 & 30.9 & 34.2 & 32.9 & 31.2\\
\rowcolor{gray!20}
\textbf{CMT-LLM (ours)} & \textbf{2025} & \textbf{21.9} & \textbf{22.8} & \textbf{30.0} & \textbf{32.9} & \textbf{29.7} & \textbf{30.4}\\
\bottomrule
\end{tabular}
\end{adjustbox}
\vspace{-5mm}
\end{table}

\begin{table*}[htbp]
\centering
\caption{WER performance comparison with different contextual ASR models on LibriMix and AMI test sets (\%).}
\vspace{-3mm}
\label{tab:performance}
\begin{adjustbox}{max width=\textwidth}
\begin{tabular}{c|c|c|c|ccc}
\toprule
\multirow{4}{*}{\textbf{Model}} & \multirow{4}{*}{\textbf{Type}} & \multirow{4}{*}{\textbf{Distractors}} & \multirow{2}{*}{\textbf{LibriMix}} & \multicolumn{3}{c}{\textbf{AMI}} \\
\cmidrule{5-7}
& & &  & \textbf{IHM-Mix} & \textbf{SDM} & \textbf{MDM} \\
\cmidrule{4-7}
& & & \textbf{WER / B-WER} & \textbf{WER / B-WER} & \textbf{WER / B-WER} & \textbf{WER / B-WER} \\
\midrule
\midrule
\multirow{1}{*}{LLM Baseline} 
& No Biasing & - & 9.2 / 25.3 & 23.5 / 45.6 & 34.2 / 51.0 & 31.2 / 49.7 \\
\cmidrule{1-7}
\hspace{1em} \multirow{3}{*}{+ ED-CEC \cite{he2023ed}} & \multirow{3}{*}{Biasing List} & + 100 & 8.4 / 15.3 & 23.1 / 34.2 & 33.6 / 35.5 & 30.8 / 32.1 \\
& & + 1,000 & 8.7 / 15.5 & 23.1 / 34.8 & 33.7 / 36.9 & 30.8 / 35.8 \\
& & + 2,000 & 8.8 / 16.0 & 23.2 / 35.1 & 33.7 / 37.7 & 30.9 / 38.7 \\
& & + 5,000 & 8.9 / 16.4 & 23.2 / 35.5 & 33.8 / 39.0 & 31.0 / 40.3 \\
\midrule
\multirow{6}{*}{CMT-LLM} 
& No Biasing & - & 9.3 / 28.7 & 23.4 / 48.7 & 33.6 / 50.8 & 31.1 / 49.5 \\
& Anti-Context & + 100 & 9.4 / 29.7 & 23.3 / 43.8 & 33.6 / 50.2 & 30.9 / 48.2 \\
\cmidrule{2-7}
& \multirow{3}{*}{Biasing List} & + 100 & 7.3 / 8.7 & 22.7 / 29.8 & 32.7 / 30.6 & 30.3 / 30.2 \\
& & + 1,000 & 7.9 / 12.5 & 22.8 / 33.6 & 32.9 / 35.0 & 30.4 / 34.6 \\
& & + 2,000 & 8.4 / 14.1 & 22.8 / 35.1 & 33.0 / 38.9 & 30.5 / 36.7 \\
& & + 5,000 & 8.4 / 15.2 & 22.9 / 36.4 & 33.1 / 42.5 & 30.6 / 38.1 \\
\cmidrule{2-7}
& GT Rare Words & - & 6.6 / 2.6 & 22.4 / 24.6 & 31.7 / 28.4 & 29.5 / 28.6 \\
\bottomrule
\end{tabular}
\vspace{-8mm}
\end{adjustbox}
\end{table*}

\subsection{Experimental Conditions}
\vspace{-2mm}
\label{sec:3.2}
\noindent \textbf{Datasets:} We evaluate the model using the LibriMix \cite{cosentino2020librimix} and AMI \cite{carletta2005ami} datasets:  
\noindent
\begin{itemize}[leftmargin=*]
\setlength{\topsep}{0pt}
\setlength{\itemsep}{0pt}
\setlength{\parsep}{0pt}
\setlength{\parskip}{0pt}
\item \noindent The \textbf{LibriMix} dataset \cite{cosentino2020librimix} combines LibriSpeech \cite{panayotov2015librispeech} speech with WHAM! noise \cite{wichern2019wham}. Following the ESPNet SOT\footnote{\href{https://github.com/espnet/espnet/tree/master/egs2/librimix/sot_asr1}{https://github.com/espnet/espnet/tree/master/egs2/librimix/sot\_asr1}}, we introduce a random delay (1.0–1.5s) between overlapping speakers, creating two-speaker mixtures. The dataset includes ~830h of speed-perturbed training data, 8.2h validation, and 7.6h test data.
\item The \textbf{AMI} dataset \cite{carletta2005ami} contains ~100h of meeting recordings with 4–5 speakers. Following the icefall SURT\footnote{\href{https://github.com/k2-fsa/icefall/tree/master/egs/ami/SURT}{https://github.com/k2-fsa/icefall/tree/master/egs/ami/SURT}}, we use three microphone settings: IHM-Mix (mixed headset mics), SDM (single distant mic), and MDM (beamformed array) \cite{anguera2007acoustic}. It includes 79.4h training, 9.7h validation, and 9.1h test data.
\end{itemize}

\begin{figure*}[t]
  \centering
  \includegraphics[width=2.1\columnwidth]{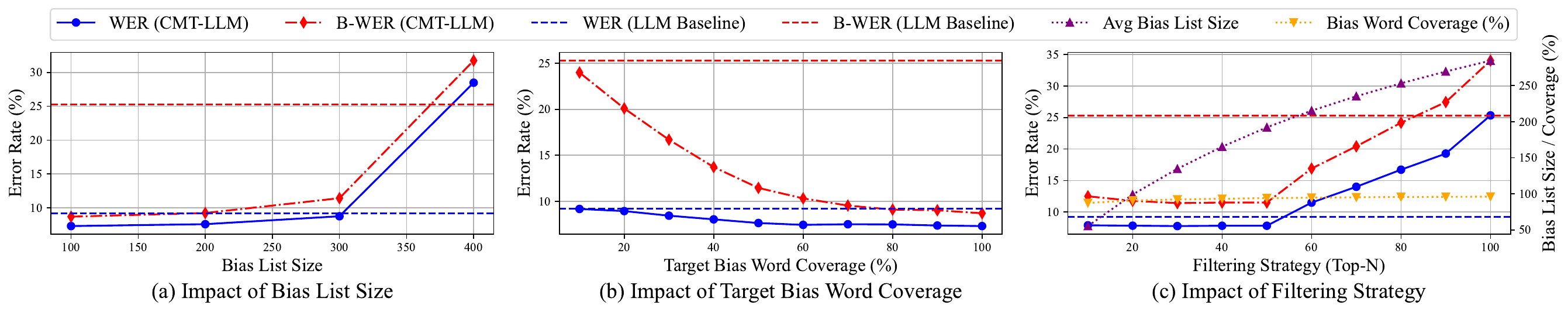}
  \vspace{-4mm}
  \caption{Impact of Biasing List Size, Coverage, and Filtering on ASR Performance.}
  \vspace{-5mm}
  \label{fig:bs}
\end{figure*}

\noindent \textbf{Biasing List Construction:} Since Librimix lacks predefined biasing lists, we follow the validated simulation method in \cite{le2021contextualized}. As Librimix is derived from Librispeech, we use the same full biasing list\footnote{\href{https://github.com/facebookresearch/fbai-speech/tree/master/is21_deep_bias}{https://github.com/facebookresearch/fbai-speech/tree/master/ \\ is21\_deep\_bias}}, containing 209.2K words. Words in this list are considered rare.  
For each utterance, we construct biasing lists by selecting words from reference transcripts that appear in the full list, adding distractors as per the experimental setup.  
To assess real-world feasibility, we construct AMI biasing lists by extracting text from lecture slides using Tesseract OCR\footnote{\href{https://github.com/tesseract-ocr/tesseract}{https://github.com/tesseract-ocr/tesseract}}, as shown in Fig. \ref{fig:pipeline} (b). Unique words are extracted, and those in the full biasing list or occurring fewer than 100 times are classified as rare. These lecture-specific lists are used for contextual correction \cite{sun2022tree}. Additionally, we merge all lecture lists into a unified AMI biasing list for large-scale experiments, selecting distractors accordingly. Note that the total number of words in the biasing lists of AMI accounts for about 1.0\% of the total vocabulary, which has a small impact on the overall WER. However, as shown in the example in Fig. \ref{fig:pipeline}, these words are mostly crucial content words, and their correct recognition is critical for understanding the utterance.

\noindent \textbf{Evaluation Metrics:} We evaluate the ASR performance using word error rate (WER) and biased WER (B-WER) \cite{le2021contextualized}, which computes the WER based on words present in the biasing list.


\vspace{-2mm}
\subsection{Results and Analysis}
\vspace{-2mm}
\label{ssec:Results and Analysis}
\noindent \textbf{Comparisons of Baselines:} Table \ref{tab:per_comparison} and Table \ref{tab:amiper_comparison} compare WER performance across different multi-talker ASR models on LibriMix and AMI datasets. CMT-LLM achieves the lowest WER in both cases, significantly outperforming the LLM Baseline and other SOTA traditional models, demonstrating its effectiveness in contexutal multi-talker ASR. 

Table \ref{tab:performance} compares different contextual ASR models. The LLM baseline, lacking a biasing list, struggles with high B-WER, highlighting the challenge of rare word recognition without explicit context. ED-CEC \cite{he2023ed}, a SOTA contextual ASR post-processing method, improves B-WER with a small biasing list (+100 distractors) but becomes less effective as the list size increases, owing to growing ambiguity from additional distractors.
In contrast, CMT-LLM, which integrates the biasing list directly into the prompt, performs significantly better with small biasing lists, surpassing both the LLM baseline and ED-CEC by a large margin. In the Anti-Context condition, where target bias words are substituted with distractors, B-WER increases significantly, highlighting the crucial role of explicit biasing. The GT Rare Words condition, which biases only the ground-truth rare words, serves as an ideal upper bound, achieving a B-WER of just 2.6\% on LibriMix, showcasing the effectiveness of optimized biasing strategies.
However, recognition performance declines when the biasing list exceeds 1,000 words, not only owing to more distractors but also reduced target biasing word coverage from filtering. Specifically, after adding 1,000, 2,000, and 5,000 distractors, coverage drops to 87.40\%, 85.07\%, and 83.07\%, respectively. Although filtering helps CMT-LLM handle large biasing lists, fewer target words negatively impact recognition. Overall, at the same target biasing word coverage, incorporating biasing information in the prompt is more effective than post-processing correction.

\noindent \textbf{Impact of Biasing List Size and Word Coverage:} Finally, we examine how biasing list size and word coverage impact recognition performance (Fig. \ref{fig:bs}).  
Without a filtering mechanism (Fig. \ref{fig:bs}(a)), increasing the biasing list size from 100 to 400 significantly raises WER and B-WER. When the list exceeds 300, WER surpasses the LLM Baseline, suggesting that an overly large list introduces interference and degrades performance.  
With the list size fixed at 100 (Fig. \ref{fig:bs}(b)), reducing word coverage from 100\% to 10\% sharply increases B-WER, indicating that low coverage weakens the biasing effect.  
By applying a filtering mechanism to an initial list containing 1,000 distractors (Fig. \ref{fig:bs}(c)), we find that if the average size of the filtered biasing list is below 200, high coverage can be maintained while minimizing interference. However, when Top-N exceeds 50, WER and B-WER spike, and at Top-60, WER surpasses the LLM Baseline, confirming that excessively large lists impair recognition.  
Future work focuses on reducing list size while maintaining high coverage to improve system robustness.

\noindent \textbf{Example of Recognizing Rare Words:} Fig. \ref{fig:pipeline}(c) provides a specific example of CMT-LLM using rare personal names extracted from slides to improve ASR accuracy, demonstrating the effectiveness of the proposed CMT-LLM.

\vspace{-2mm}
\section{Conclusion}
This paper for the first time introduces an LLM-based SOT method CMT-LLM for multi-talker contextual ASR. Leveraging its strong decoding capabilities, deep comprehension of long-range context, and cross-speaker modeling ability, LLM excels in handling complex multi-talker speech scenarios.  In addition, CMT-LLM effectively incorporates contextual information through prompt learning, leading to a significant improvement in recognizing rare words. Specifically, for large biasing lists containing thousands of words, a coarse decoding-based filtering algorithm is applied to substantially reduce the list size, preserving only the 10 most relevant words for each remaining word and integrating them into the prompt input. This further enhances the contextual biasing effect even with 5,000 distractors.  
Experimental results demonstrate that our approach achieves state-of-the-art recognition performance on both the simulated LibriMix dataset and the real-world AMI dataset, significantly outperforming existing conventional methods.

\section{Acknowledgments}
This work was partly supported by JST CREST Grant Number JPMJCR22D1, Japan.

\bibliographystyle{IEEEtran}
\bibliography{mybib}

\end{document}